\documentclass[referee,usenatbib]{mn2e}
\usepackage{graphicx,subfigure}
\usepackage{bm}
\usepackage{longtable}
\usepackage{amsmath,amssymb,graphicx}
\usepackage{rotating}
\usepackage{color}
\newcommand{\g}{\gamma}
\newcommand{\gm}{\gamma_{\rm m}}

\graphicspath{{./}{Figures/}}
\title[Compton Scattering of Self-Absorbed Synchrotron Emission]
{Compton Scattering of Self-Absorbed Synchrotron Emission}
\author[Gao, Lei, Wu, \& Zhang  ]{He Gao$^{1}$, Wei-Hua Lei$^{2,1}$, Xue-Feng Wu$^{3,4}$ and Bing Zhang$^{1,5}$\\
1. Department of Physics \& Astronomy, University of Nevada, Las Vegas, NV 89154-4002, USA.\\
gaohe@physics.unlv.edu, zhang@physics.unlv.edu\\
2. School of Physics, Huazhong University of Science and Technology,
Wuhan, 430074, China.\\
3. Purple Mountain Observatory, Chinese Academy of Sciences,
Nanjing, 210008, China.\\
4. Chinese Center for Antarctic Astronomy, Nanjing, 210008, China.\\
5. Kavli Insitute for Astronomy and Astrophysics and Department of Astronomy, Peking University,
Beijing 100871, China.}
\begin{document}
\date{Accepted.....; Received .....}

\pagerange{\pageref{firstpage}--\pageref{lastpage}} \pubyear{}

\maketitle

\label{firstpage}

\begin{abstract}
Synchrotron self-Compton (SSC) scattering is an important emission
mechanism in many astronomical sources, such as gamma-ray bursts
(GRBs) and active galactic nuclei (AGNs). We give a complete
presentation of the analytical approximations for the Compton
scattering of synchrotron emission with both weak and strong
synchrotron self-absorption. All possible orders of the
characteristic synchrotron spectral breaks ($\nu_{\rm a}$, $\nu_{\rm
m}$, and $\nu_{\rm c}$) are studied. In the weak self-absorption
regime, i.e.,  $\nu_{\rm a} < \nu_c$, the electron energy
distribution is not modified by the self-absorption process. The
shape of the SSC component broadly resembles that of synchrotron,
but with the following features: The SSC flux increases linearly
with frequency up to the SSC break frequency corresponding to the
self-absorption frequency $\nu_{\rm a}$; and the presence of a
logarithmic term in the high-frequency range of the SSC spectra
makes it harder than the power-law approximation. In the strong
absorption regime, i.e. $\nu_{\rm a} > \nu_{\rm c}$, heating of low
energy electrons due to synchrotron absorption leads to pile-up of
electrons, and form a thermal component besides the broken power-law
component. This leads to two-component (thermal + non-thermal)
spectra for both the synchrotron and SSC spectral components. For
$\nu_{\rm c} < \nu_{\rm a} < \nu_{\rm m}$, the spectrum is thermal
(non-thermal) -dominated if $\nu_a > \sqrt{\nu_m \nu_c}$ ($\nu_a <
\sqrt{\nu_m \nu_c}$). Similar to the weak-absorption regime, the SSC
spectral component is broader than the simple broken power law
approximation. We derive the critical condition for strong
absorption (electron pile-up), and discuss a case of GRB reverse
shock emission in a wind medium, which invokes $\nu_{\rm a}
> {\rm max} (\nu_{\rm m}, \nu_{\rm c})$.
\end{abstract}



\section{Introduction}

Astrophysical sources powered by synchrotron radiation should have a
synchrotron self-Compton (SSC) scattering component. The same
electrons that radiate synchrotron photons would scatter these
synchrotron seed photons to high energies, forming a distinct
spectral component. The SSC mechanism has been invoked to account
for the observed high energy emission in many astrophysical sources,
such as gamma-ray bursts (GRBs)
\citep[e.g.][]{Meszaros94,Wei98,Dermer00,Panaitescu00,Sari01,Zhang01,Wang01,Wu04,
Zou09} and active galactic nuclei (AGNs)
\citep[e.g.][]{Ghisellini98,Chiang02,Zhang12}.

SSC is a complex process. The flux at each observed frequency
includes the contributions from electrons in a wide range of energies,
which scatter seed photons in a wide range of frequencies. Therefore,
a precise description of the SSC spectrum invokes a complex
convolution of the seed photon spectrum and electron energy
distribution, which requires numerical calculations. However, for a
synchrotron source with shock-accelerated electrons, the injected
electron spectrum is usually assumed to be a simple power-law
function, the corresponding electron energy distribution and seed
synchrotron spectrum thus have simple patterns. Some analytical
approximations for the SSC spectrum can be then made if Compton
scattering is in the Thomson regime.

Besides the injected electron spectrum, two other factors are
essential to define the shape of the final electron energy
distribution in a synchrotron source: radiation cooling and
self-absorption heating. There are three characteristic synchrotron
frequencies in the spectrum: the minimum injection frequency
($\nu_{\rm m}$), the cooling frequency ($\nu_c$), and the
self-absorption frequency ($\nu_{\rm a}$). When $\nu_{\rm a} <
\nu_{\rm c}$, the heating effect due to self-absorption is not
important in modifying the electron energy spectrum.  For a
continuous injection of a power-law electron spectrum, the final
electron energy distribution is a broken power law. The seed
synchrotron spectrum for SSC is characterized by a multi-segment
broken power law, separated by $\nu_m$, $\nu_c$, and $\nu_a$.
Different ordering of the three characteristic frequencies leads to
different shapes of the seed synchrotron spectrum. In the
literature, usually $\nu_{\rm a} < {\rm min}(\nu_{\rm m}, \nu_{\rm
c})$ is assumed. \cite{Sari01} have derived the approximated
expressions of the SSC spectrum in the $\nu_{\rm a} < \nu_{\rm m} <
\nu_{\rm c}$ and $\nu_{\rm a} < \nu_{\rm c} < \nu_{\rm m}$ regimes,
respectively\footnote{Assuming weak self-absorption, Gou et al.
(2007) derived analytical approximations of the SSC component for
several other spectral regimes.}.

When $\nu_{\rm a}>\nu_{\rm c}$, synchrotron self-absorption
becomes an important heating source for the low-energy electrons.
Consequently, the electrons are dominated by a quasi-thermal component
until a ``transition'' Lorentz factor $\gamma_{\rm t}$, above which
the electrons are no longer affected by the self-absorption heating
and keep the normal power law distribution
\citep{Ghisellini88,Ghisellini91,Ghisellini98a}. For these strong
absorption cases, a thermal peak due to pile-up electrons would appear
around $\nu_{\rm a}$ in the synchrotron spectrum \citep{Kobayashi04},
which would also result in some new features in the SSC spectrum.

In this paper, we extend the analysis of \cite{Sari01} and present
the full analytical approximated expressions of the SSC spectrum in
all six possible cases of $\nu_{\rm a}$, $\nu_{\rm m}$, $\nu_{\rm c}$ ordering. In
Section 2, three weak synchrotron self-absorption cases ($\nu_{\rm a}<\nu_{\rm c}$)
are discussed. In Section 3, we focus on the strong synchrotron
self-absorption regime ($\nu_{\rm a}>\nu_{\rm c}$), where synchrotron
self-absorption significantly affects the electron energy distribution.
By adopting a simplified prescription of the pile-up electron distribution,
we derive the expressions of both synchrotron and SSC spectral
components. All the expressions in this work are valid in
the Thomson regime, so that the Klein-Nishina correction
effect \citep[e.g.][]{Rees67,Nakar09}
is not important in the first order SSC component. We also limit our
treatment to the first-order SSC, and assume that the higher-order SSC
components \citep[e.g.][]{Kobayashi07,Piran09} are suppressed by the
Klein-Nishina effect. Such an assumption is usually valid for most
problems. In order to make a simple analytical treatment, we have applied
a simplified approximation for the synchrotron spectra, and adopted the
simplification that the inverse Compton scattering of mono-energetic
electrons off mono-energetic seed photons is also mono-energetic
\citep{Sari01}. This would not significantly deteriorate precision of
the analysis, while making it much simpler.

\section{Weak Synchrotron Self-absorption Cases}

In the single scattering regime, the inverse Compton volume emissivity for a
power-law distribution of electrons is \citep{ryl79,Sari01}
\begin{eqnarray}
\label{jc} j^{IC}_{\nu} = 3 \sigma_T \int_{\gm}^{\infty}{d \g N(\g)
\int_0^1{d x\,g(x) \tilde{f}_{\nu_s}(x)}},
\end{eqnarray}
where $x \equiv \nu/4 \g^2 \nu_s$ (an angle-dependent parameter),
$\tilde{f}_{\nu_s}$ is the incident specific flux in the shock front,
$\sigma_T$ is Thomson scattering cross section, and $g(x) = 1+x+2 x
\ln{x}-2 x^2$ takes care of the angular dependence of the scattering
cross section in
the limit of $\g \gg 1$ \citep{blg70}. One can approximate $g(x)= 1$
for $0<x<x_0$ to simplify the integration, which would yield a correct
behavior for $x\ll 1$ (Sari \& Esin 2001). With such a simplification,
the SSC spectrum is given by (Sari \& Esin 2001),
\begin{eqnarray}
\label{fc} f^{\rm{IC}}_{\nu} = R \sigma_T \int_{\gm}^{\infty}{d \g
N(\g) \int_0^{x_0}{d x\, f_{\nu_{\rm s}}(x)}},
\end{eqnarray}
where $f_{\nu_{\rm s}}(x)$ is the synchrotron flux,  $R$ is the
co-moving size of the emission region, and the value of the
parameter $x_0$ is set by ensuring energy conservation, i.e.
$\int_0^1{x\, g(x) d x} = \int_0^{x_0}{x\, d x}$.

When $\nu_{\rm a}<\nu_{\rm c}$, in the slow cooling regime
($\gamma_{\rm m} < \gamma_{\rm c}$), the electron energy
distribution is
\begin{eqnarray}
\label{Ngam1} N(\gamma) = \left\{ \begin{array}{ll}
n(p-1)\gamma_{\rm m}^{p-1}\gamma^{-p}, &  \gamma_{\rm m} \leq \gamma \leq \gamma_{\rm c}, \\
n(p-1)\gamma_{\rm m}^{p-1}\gamma_{\rm c}\gamma^{-p-1}, &  \gamma > \gamma_{\rm c}.
\end{array} \right.
\end{eqnarray}
Here $\gamma_{\rm m}$ is the minimum Lorentz factor of the injected
electrons, and $p$ is electron spectral index. Cooling is efficient
for electrons with Lorentz factor above the critical value
$\gamma_{\rm c}$. Notice that Eq.\ref{Ngam1} is only valid for
$p>1$.

In the fast cooling regime ($\gamma_{\rm c} < \gamma_{\rm m}$), the electron
energy distribution is\footnote{This is valid only in the deep fast cooling regime.
For a non-steady state with not too deep fast cooling, the electron spectrum can
be harder than -2 \citep{Uhm13}.}
\begin{eqnarray}
\label{fnu1} N(\gamma) = \left\{ \begin{array}{ll}
n \gamma_{\rm c} \gamma^{-2}, &  \gamma_{\rm c} \leq \gamma \leq \gamma_{\rm m}, \\
n \gamma_{\rm m}^{p-1}\gamma_{\rm c} \gamma^{-p-1}, &  \gamma >
\gamma_{\rm m}.
\end{array} \right.
\end{eqnarray}
In this regime, all the injected electrons are able to cool on
the dynamical timescale. Therefore, there is a population of
electrons with Lorentz factor below the injection minimum Lorentz
factor $\gamma_{\rm m}$.

The seed synchrotron spectrum $f_{\nu_{\rm s}}$ has spectral beaks at
$\nu_{\rm a}$, $\nu_{\rm m}$ and $\nu_{\rm c}$, where $\nu_{\rm a}$
is the self-absorption frequency, below which the system becomes
optically thick, and $\nu_{\rm m}$ and $\nu_{\rm c}$ are
the characteristic synchrotron frequencies for the electrons with
Lorentz factors $\gamma_{\rm m}$ and $\gamma_{\rm c}$, respectively.

As shown in \cite{Sari01}, the critical frequencies in the SSC component
are defined by different combination of $\gamma_{\rm a}, \gamma_{\rm m},
\gamma_{\rm c}$ and $\nu_{\rm a},\nu_{\rm m}, \nu_{\rm c}$. For convenience, we use a new
notation in this paper
\begin{eqnarray}
\nu_{ij}^{\rm{IC}}=4 \gamma_i^2 \nu_j
x_0,~~~~~~~~~~~~~~~~~~~~~i,j={a,c,m}. \label{nus1}
\end{eqnarray}
The physical meaning is the characteristic upscattered frequency for
mono-energetic electrons with Lorentz factor $\gamma_i$ scattering off
mono-energetic photons with frequency $\nu_j$.

\subsection{Case I: $\nu_{\rm a} < \nu_{\rm m} < \nu_{\rm c}$}

This case has been studied by \cite{Sari01}. The synchrotron
spectrum reads\footnote{Hereafter, the synchrotron spectra are
denoted as $f_\nu(\nu)$ for simple presentation. Notice that when
they are taken as seed spectrum, one should consider them as
$f_{\nu_{\rm s}}(\nu_{\rm s})$ and apply equation (\ref{fc}) to calculate the
SSC spectra.}
\begin{eqnarray}
\label{fnu1} f_{\nu} = \left\{ \begin{array}{ll}
f_{\rm{max}}\left(\frac{\nu_{\rm a}}{\nu_{\rm m}}\right)^{\frac{1}{3}}
\left(\frac{\nu}{\nu_{\rm a}}\right)^{2}, &
\nu \leq \nu_{\rm a}; \\
f_{\rm{max}}\left(\frac{\nu}{\nu_{\rm m}}\right)^{\frac{1}{3}}, &
\nu_{\rm a} < \nu \leq \nu_{\rm m}; \\
f_{\rm{max}}\left(\frac{\nu}{\nu_{\rm m}}\right)^{\frac{1-p}{2}}, &
\nu_{\rm m} < \nu \leq \nu_{\rm c}; \\
f_{\rm{max}}\left(\frac{\nu_{\rm c}}{\nu_{\rm m}}\right)^{\frac{1-p}{2}}
\left(\frac{\nu}{\nu_{\rm c}}\right)^{-\frac{p}{2}}, &
\nu > \nu_{\rm c}, \\
\end{array} \right.
\end{eqnarray}
where $f_{\rm{max}} = f_{\nu} (\nu_{\rm m})$ is the peak flux density of
the synchrotron component, which is taken as a constant.
Substituting this seed photon spectrum into equation (\ref{fc}), the
inner integral reads \citep{Sari01}
\begin{eqnarray}
\label{int1} I = \left\{ \begin{array}{ll} I_1 \simeq \frac{5}{2}
f_{\rm{max}} x_0 \left(\frac{\nu_{\rm a}}{\nu_{\rm m}}\right)^{\frac{1}{3}}
\left(\frac{\nu}{4 \g^2 \nu_{\rm a} x_0}\right), & \nu < 4 \g^2 \nu_{\rm a} x_0 \\
I_2 \simeq \frac{3}{2} f_{\rm{max}} x_0 \left(\frac{\nu}{4 \g^2 \nu_{\rm m}
x_0}\right)^{\frac{1}{3}}, &
4 \g^2 \nu_{\rm a} x_0 < \nu < 4 \g^2 \nu_{\rm m} x_0 \\
I_3 \simeq \frac{2}{(p+1)} f_{\rm{max}} x_0 \left(\frac{\nu}{4 \g^2 \nu_{\rm m}
x_0}\right)^{\frac{1-p}{2}}, &
4 \g^2 \nu_{\rm m} x_0 < \nu < 4 \g^2 \nu_{\rm c} x_0 \\
I_4 \simeq \frac{2}{(p+2)} f_{\rm{max}} x_0
\left(\frac{\nu_{\rm c}}{\nu_{\rm m}}\right)^{\frac{1-p}{2}} \left(\frac{\nu}{4
\g^2 \nu_{\rm c} x_0}\right)^{-\frac{p}{2}}, & \nu > 4 \g^2 \nu_{\rm c} x_0.
\end{array} \right.
\end{eqnarray}
Similar to \cite{Sari01}, only the leading order of $\nu$ and zeroth
order of $\nu_{\rm a}/\nu_{\rm m}$ and $\nu_{\rm m}/\nu_{\rm c}$ are shown.
However, we note
that higher order small terms are needed to derive the following SSC
spectrum (\ref{fc2-1}) through integrating the outer integral of
equation (\ref{fc}).

After integration, $f^{\rm{IC}}_{\nu}$ is very complex.
Keeping only the dominant terms, one gets the analytical approximation
\begin{eqnarray}
\label{fc2-1} f_{\nu}^{\rm{IC}} &\simeq& R \sigma_T n f_{\rm{max}} x_0 \\
\nonumber &\times&\left\{ \begin{array}{ll}

\frac{5}{2} \frac{(p-1)}{(p+1)}
\left(\frac{\nu_{\rm a}}{\nu_{\rm m}}\right)^{\frac{1}{3}}
\left(\frac{\nu}{\nu_{\rm ma}^{\rm{IC}}}\right), &
\nu < \nu_{\rm ma}^{\rm{IC}}; \\

\frac{3}{2} \frac{(p-1)}{(p-1/3)}
\left(\frac{\nu}{\nu_{\rm mm}^{\rm{IC}}}\right)^{\frac{1}{3}}, &
\nu_{\rm ma}^{\rm{IC}} < \nu < \nu_{\rm mm}^{\rm{IC}}; \\

\frac{(p-1)}{(p+1)}
\left(\frac{\nu}{\nu_{\rm mm}^{\rm{IC}}}\right)^{\frac{1-p}{2}}
\left[\frac{4 (p+1/3)}{(p+1)(p-1/3)} +
\ln{\left(\frac{\nu}{\nu_{\rm mm}^{\rm{IC}}}\right)}\right], &
\nu_{\rm mm}^{\rm{IC}} < \nu < \nu_{\rm mc}^{\rm{IC}}; \\

\frac{(p-1)}{(p+1)}
\left(\frac{\nu}{\nu_{\rm mm}^{\rm{IC}}}\right)^{\frac{1-p}{2}}
\left[\frac{2(2 p+3)}{(p+2)} - \frac{2}{(p+1)(p+2)} +
\ln{\left(\frac {\nu_{\rm cc}^{\rm{IC}}}{\nu}\right)}\right], &
\nu_{\rm mc}^{\rm{IC}} < \nu < \nu_{\rm cc}^{\rm{IC}}; \\

\frac{(p-1)}{(p+1)}
\left(\frac{\nu}{\nu_{\rm mm}^{\rm{IC}}}\right)^{-\frac{p}{2}}
\left(\frac{\nu_{\rm c}}{\nu_{\rm m}}\right) \left[\frac{2(2p+3)}{(p+2)} -
\frac{2}{(p+2)^2} + \frac{(p+1)}{(p+2)}
\ln{\left(\frac{\nu}{\nu_{\rm cc}^{\rm{IC}}}\right)}\right], & \nu >
\nu_{\rm cc}^{\rm{IC}}.
\end{array} \right.
\end{eqnarray}
Notice that \cite{Sari01} presented an opposite sign for
the term $\frac{2}{(p+2)^2}$ in the last segment, which might
be a typo in that paper.

\begin{figure}
\begin{minipage}[b]{1.0\textwidth}
\centering
\includegraphics[height=2.8in,width=5in]{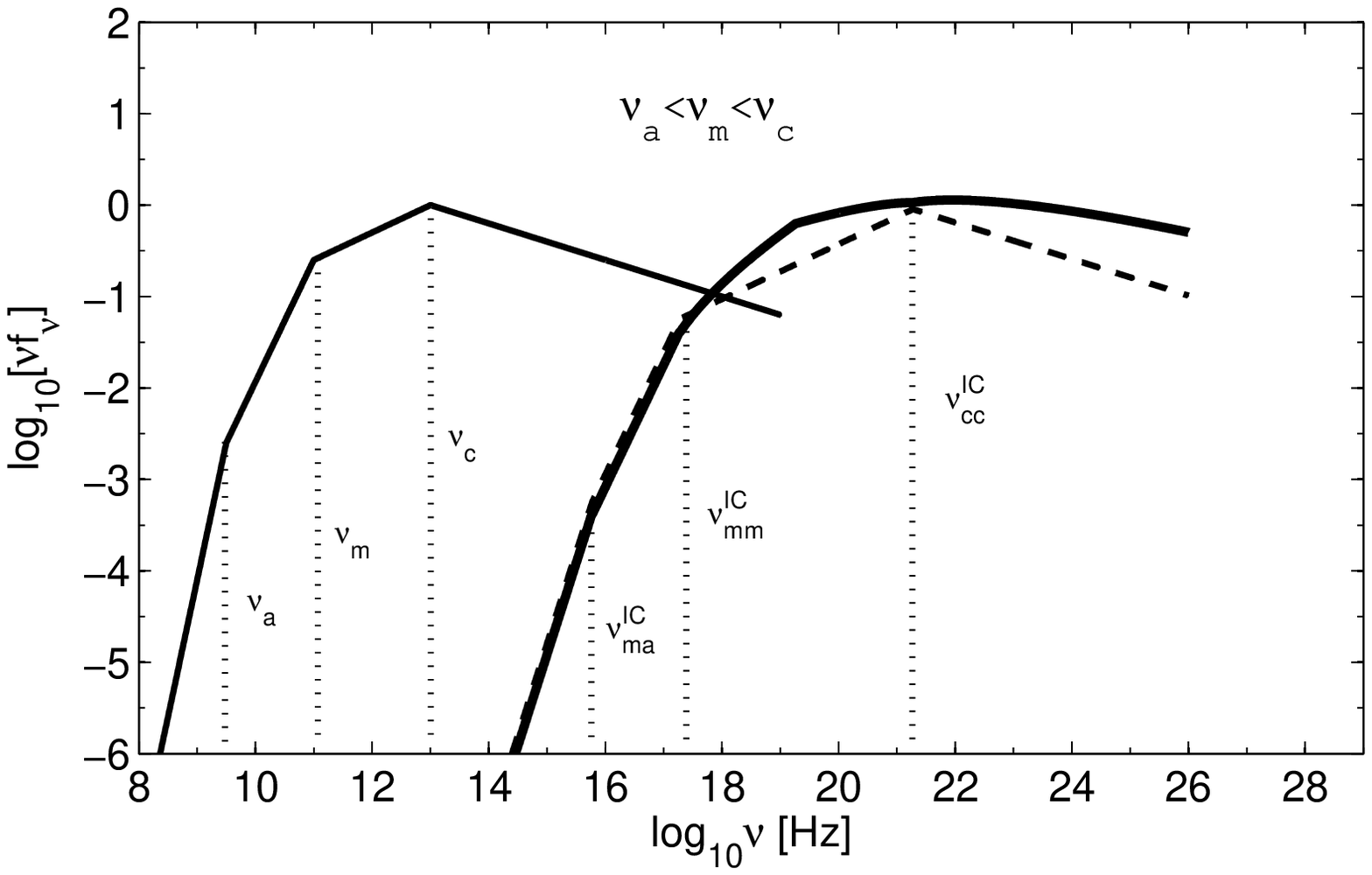}
\end{minipage} \\%
\begin{minipage}[b]{1.0\textwidth}
\centering
\includegraphics[height=2.8in,width=5in]{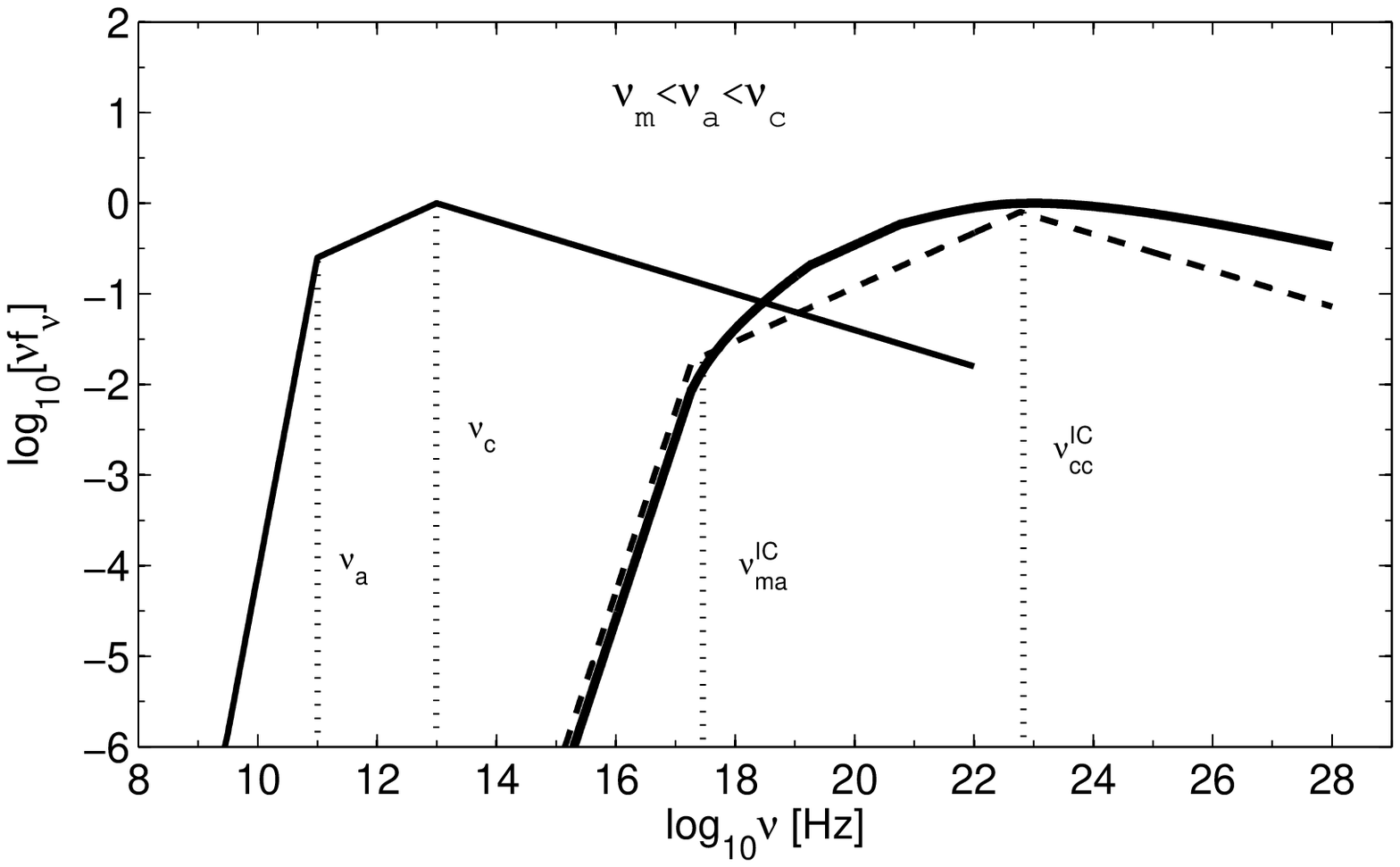}
\end{minipage} \\%
\begin{minipage}[b]{1.0\textwidth}
\centering
\includegraphics[height=2.8in,width=5in]{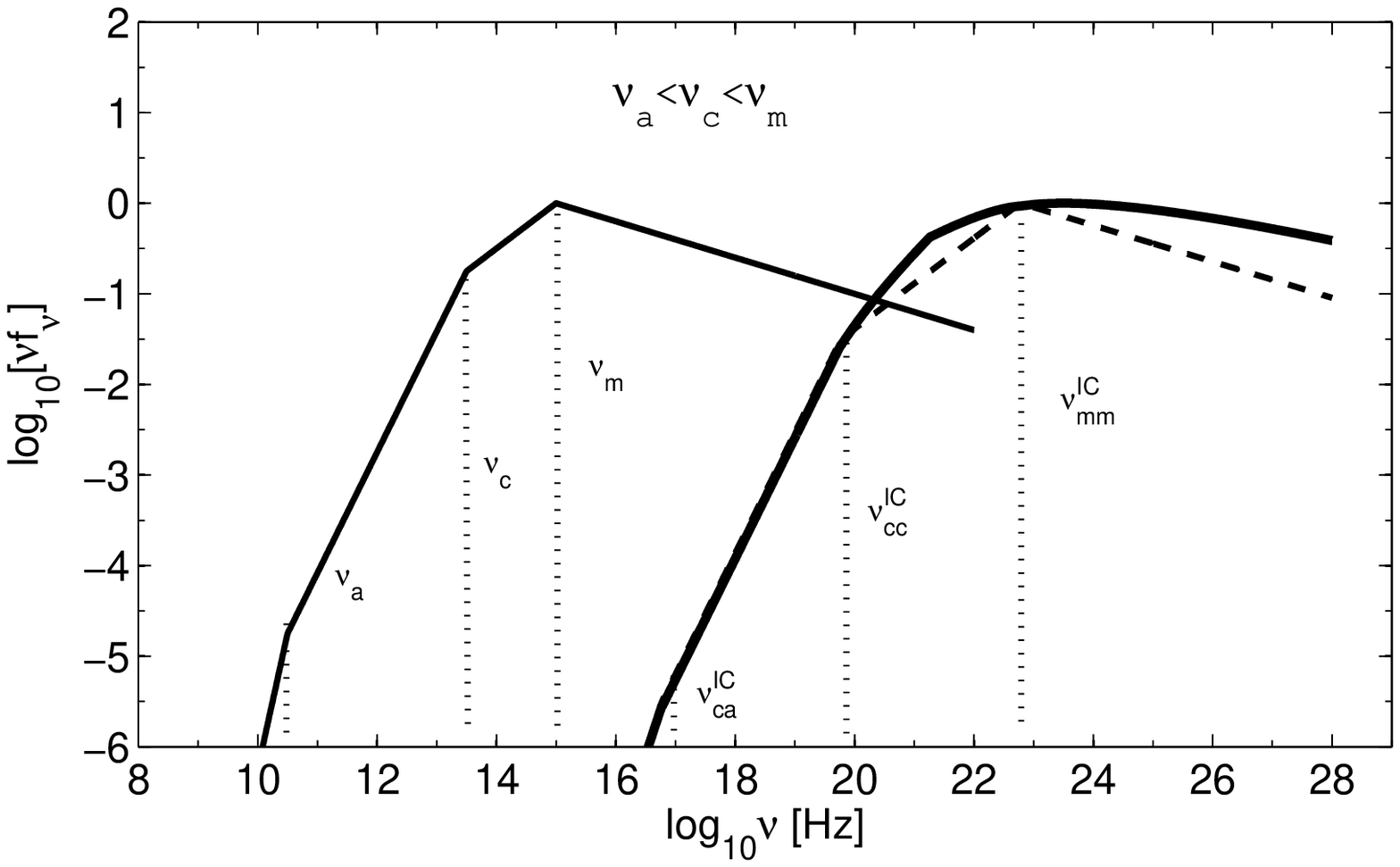}
\end{minipage}%
              \caption{Total synchrotron $+$ SSC spectra for weak synchrotron reabsorption cases ($\nu_{\rm a}<\nu_{\rm c}$).
               The top panel is for $\nu_{\rm a} <
              \nu_{\rm m} <\nu_{\rm c}$ case; the middle panel is for $\nu_{\rm m} <
              \nu_{\rm a} <\nu_{\rm c}$ case; and the bottom panel is for $\nu_{\rm a} <
              \nu_{\rm c} <\nu_{\rm m}$ case. The thin solid line is
       synchrotron component. The thick solid line in the SSC component is drawn using the
analytical approximations, while the dashed lines are the broken
power-law approximation for comparison. In all the cases, the $\nu F_\nu$ peaks
for both the synchrotron and the SSC components are normalized to unity.}
           \label{Fig1}
            \end{figure}

The normalized synchrotron + SSC spectra for this and other two weak
self-absorption cases are presented in Figure \ref{Fig1}. We note that
these analytical expressions are not continuous around the breaks
because of dropping the small order terms (see also Sari \& Esin
2001), but the mis-match is small. When plotting the SSC curve in
Figure \ref{Fig1}, we have used the analytical approximations,
but added back some smaller order terms to remove the discontinuity.

\subsection{Case II: $\nu_{\rm m} < \nu_{\rm a} < \nu_{\rm c}$}

The synchrotron photons spectrum reads
\begin{eqnarray}
\label{fnu2} f_{\nu} = \left\{ \begin{array}{ll}
f_{\rm{max}}\left(\frac{\nu_{\rm m}}{\nu_{\rm a}}\right)^{\frac{p+4}{2}}
\left(\frac{\nu}{\nu_{\rm m}}\right)^{2}, &
\nu \leq \nu_{\rm m}; \\
f_{\rm{max}}\left(\frac{\nu_{\rm a}}{\nu_{\rm m}}\right)^{\frac{1-p}{2}}\left(\frac{\nu}{\nu_{\rm a}}\right)^{\frac{5}{2}},
&
\nu_{\rm m} < \nu \leq \nu_{\rm a}; \\
f_{\rm{max}}\left(\frac{\nu}{\nu_{\rm m}}\right)^{\frac{1-p}{2}}, &
\nu_{\rm a} < \nu \leq \nu_{\rm c}; \\
f_{\rm{max}}\left(\frac{\nu_{\rm c}}{\nu_{\rm m}}\right)^{\frac{1-p}{2}}
\left(\frac{\nu}{\nu_{\rm c}}\right)^{-\frac{p}{2}}, &
\nu > \nu_{\rm c}; \\
\end{array} \right.
\end{eqnarray}

Evaluating the inner integral in equation (\ref{fc}), we obtain
\begin{eqnarray}
\label{int2} I = \left\{ \begin{array}{ll} I_1 \simeq
\frac{2(p+4)}{3(p+1)} f_{\rm{max}} x_0
\left(\frac{\nu_{\rm m}}{\nu_{\rm a}}\right)^{\frac{p+1}{2}}
\frac{\nu}{4 \g^2 \nu_{\rm m} x_0}, & \nu < 4 \g^2 \nu_{\rm a} x_0 \\
I_2 \simeq \frac{2}{p+1} f_{\rm{max}} x_0 \left(\frac{\nu}{4 \g^2 \nu_{\rm m}
x_0}\right)^{\frac{1-p}{2}}, &
4 \g^2 \nu_{\rm a} x_0 < \nu < 4 \g^2 \nu_{\rm c} x_0 \\
I_3 \simeq \frac{2}{(p+2)} f_{\rm{max}} x_0
\left(\frac{\nu_{\rm c}}{\nu_{\rm m}}\right)^{\frac{1}{2}}\left(\frac{\nu}{4
\g^2 \nu_{\rm m} x_0}\right)^{-\frac{p}{2}}, &
\nu > 4 \g^2 \nu_{\rm c} x_0 \\
\end{array} \right.
\end{eqnarray}

An interesting feature of this result is that $I_1$ is linear with
$\nu$ all the way to $\nu = 4 \g^2 \nu_{\rm a} x_0$, indicating that a
break corresponding to the break in the synchrotron spectrum at
$\nu_{\rm m}$ does not show up in the SSC spectrum for monoenergetic
electron scattering. When $\nu > 4 \g^2 \nu_{\rm a} x_0$, the SSC spectrum
follows the same frequency dependence as the corresponding seed
synchrotron spectrum.

After second integration, we get the analytical approximation in
this regime:
\begin{eqnarray}
\label{fc2} f_{\nu}^{\rm{IC}} &\simeq& R \sigma_T n f_{\rm{max}} x_0 \\
\nonumber &\times&\left\{ \begin{array}{ll}

\frac{2(p+4)(p-1)}{3(p+1)^2}
\left(\frac{\nu_{\rm m}}{\nu_{\rm a}}\right)^{\frac{p+1}{2}}
\left(\frac{\nu}{\nu_{\rm mm}^{\rm{IC}}}\right), &
\nu < \nu_{\rm ma}^{\rm{IC}}; \\

\frac{(p-1)}{(p+1)}
\left(\frac{\nu}{\nu_{\rm mm}^{\rm{IC}}}\right)^{\frac{1-p}{2}}
\left[\frac{ 2(2p+5)}{(p+1)(p+4)} +
\ln{\left(\frac{\nu}{\nu_{\rm ma}^{\rm{IC}}}\right)}\right], &
\nu_{\rm ma}^{\rm{IC}} < \nu < \nu_{\rm mc}^{\rm{IC}}; \\

\frac{(p-1)}{(p+1)}
\left(\frac{\nu}{\nu_{\rm mm}^{\rm{IC}}}\right)^{\frac{1-p}{2}} \left[2
+ \frac{2}{p+4} + \ln{\left(\frac {\nu_{\rm c}}{\nu_{\rm a}}\right)}\right], &
\nu_{\rm mc}^{\rm{IC}} < \nu < \nu_{\rm ca}^{\rm{IC}}; \\

\frac{(p-1)}{(p+1)}
\left(\frac{\nu}{\nu_{\rm mm}^{\rm{IC}}}\right)^{\frac{1-p}{2}}
\left[\frac{2(2 p+1)}{(p+1)} + \ln{\left(\frac
{\nu_{\rm cc}^{\rm{IC}}}{\nu}\right)}\right], &
\nu_{\rm ca}^{\rm{IC}} < \nu < \nu_{\rm cc}^{\rm{IC}}; \\

\frac{(p-1)}{(p+2)}
\left(\frac{\nu_{\rm c}}{\nu_{\rm m}}\right)\left(\frac{\nu}{\nu_{\rm mm}^{\rm{IC}}}\right)^{-\frac{p}{2}}
\left[\frac{2(2p+5)}{(p+2)}+
\ln{\left(\frac{\nu}{\nu_{\rm cc}^{\rm{IC}}}\right)}\right], & \nu >
\nu_{\rm cc}^{\rm{IC}}.
\end{array} \right.
\end{eqnarray}

Similar to the $I$ result, there is no spectral break around
$\nu_{\rm mm}^{\rm{IC}}$. Another comment is that the logarithmic terms
make the SSC spectrum harder than the simple broken power-law
approximation above the $\nu F_\nu$ peak frequency. At high
frequencies, the simple broken power-law approximation may not be
adequate to represent the true SSC spectrum.

\subsection{Case III: $\nu_{\rm a} < \nu_{\rm c} < \nu_{\rm m}$}

This case was also studied by \cite{Sari01}. The seed synchrotron
spectrum reads
\begin{eqnarray}
\label{fnu1fc} f_{\nu} = \left\{ \begin{array}{ll}
f_{\rm{max}}\left(\frac{\nu_{\rm a}}{\nu_{\rm c}}\right)^{\frac{1}{3}}
\left(\frac{\nu}{\nu_{\rm a}}\right)^{2}, &
\nu \leq \nu_{\rm a}; \\
f_{\rm{max}}\left(\frac{\nu}{\nu_{\rm c}}\right)^{\frac{1}{3}}, &
\nu_{\rm a} < \nu \leq \nu_{\rm c}; \\
f_{\rm{max}}\left(\frac{\nu}{\nu_{\rm c}}\right)^{-\frac{1}{2}}, &
\nu_{\rm c} < \nu \leq \nu_{\rm m}; \\
f_{\rm{max}}\left(\frac{\nu_{\rm c}}{\nu_{\rm m}}\right)^{\frac{1}{2}}
\left(\frac{\nu}{\nu_{\rm m}}\right)^{-\frac{p}{2}}, &
\nu > \nu_{\rm m}; \\
\end{array} \right.
\end{eqnarray}
This gives
\begin{eqnarray}
\label{int1fc} I = \left\{ \begin{array}{ll} I_1 \simeq \frac{5}{2}
f_{\rm{max}} x_0 \left(\frac{\nu_{\rm a}}{\nu_{\rm c}}\right)^{\frac{1}{3}}
\left(\frac{\nu}{4 \g^2 \nu_{\rm a} x_0}\right), & \nu < 4 \g^2 \nu_{\rm a} x_0 \\
I_2 \simeq \frac{3}{2} f_{\rm{max}} x_0 \left(\frac{\nu}{4 \g^2
\nu_{\rm c} x_0}\right)^{\frac{1}{3}}, &
4 \g^2 \nu_{\rm a} x_0 < \nu < 4 \g^2 \nu_{\rm c} x_0 \\
I_3 \simeq \frac{2}{3} f_{\rm{max}} x_0 \left(\frac{\nu}{4 \g^2
\nu_{\rm c} x_0}\right)^{-\frac{1}{2}}, &
4 \g^2 \nu_{\rm c} x_0 < \nu < 4 \g^2 \nu_{\rm m} x_0 \\
I_4 \simeq \frac{2}{(p+2)} f_{\rm{max}} x_0
\left(\frac{\nu_{\rm c}}{\nu_{\rm m}}\right)^{\frac{1}{2}} \left(\frac{\nu}{4
\g^2 \nu_{\rm m} x_0}\right)^{-\frac{p}{2}}, & \nu > 4 \g^2 \nu_{\rm m} x_0
\end{array} \right.
\end{eqnarray}
and the final SSC spectrum
\begin{eqnarray}
\label{fc1fc} f_{\nu}^{\rm{IC}} &\simeq& R \sigma_T n f_{\rm{max}} x_0 \\
\nonumber &\times&\left\{ \begin{array}{ll}

\frac{5}{6} \left(\frac{\nu_{\rm a}}{\nu_{\rm c}}\right)^{\frac{1}{3}}
\left(\frac{\nu}{\nu_{\rm ca}^{\rm{IC}}}\right), &
\nu < \nu_{\rm ca}^{\rm{IC}}; \\

\frac{9}{10}
\left(\frac{\nu}{\nu_{\rm cc}^{\rm{IC}}}\right)^{\frac{1}{3}}, &
\nu_{\rm ca}^{\rm{IC}} < \nu < \nu_{\rm cc}^{\rm{IC}}; \\

\frac{1}{3}
\left(\frac{\nu}{\nu_{\rm cc}^{\rm{IC}}}\right)^{-\frac{1}{2}}
\left[\frac{28}{15} +
\ln{\left(\frac{\nu}{\nu_{\rm cc}^{\rm{IC}}}\right)}\right], &
\nu_{\rm cc}^{\rm{IC}} < \nu < \nu_{\rm cm}^{\rm{IC}}; \\

\frac{1}{3}
\left(\frac{\nu}{\nu_{\rm cc}^{\rm{IC}}}\right)^{-\frac{1}{2}}
\left[\frac{2(p+5)}{(p+2)(p-1)} - \frac{2 (p-1)}{3 (p+2)} +
\ln{\left(\frac{\nu_{\rm mm}^{\rm{IC}}}{\nu}\right)}\right], &
\nu_{\rm cm}^{\rm{IC}} < \nu < \nu_{\rm mm}^{\rm{IC}}; \\

\frac{1}{(p+2)}
\left(\frac{\nu_{\rm c}}{\nu_{\rm m}}\right)\left(\frac{\nu}{\nu_{\rm mm}^{\rm{IC}}}\right)^{-\frac{p}{2}}
\left[\frac{2}{3} \frac{(p+5)}{(p-1)} -
\frac{2}{3}\frac{(p-1)}{(p+2)} +
\ln{\left(\frac{\nu}{\nu_{\rm mm}^{\rm{IC}}}\right)}\right], & \nu >
\nu_{\rm mm}^{\rm{IC}}.
\end{array} \right.
\end{eqnarray}
We note that \cite{Sari01} has an opposite sign in the term
$\ln{\left(\frac{\nu}{\nu_{\rm cc}^{\rm{IC}}}\right)}$ in the third
segment, which might be another typo in that paper.

We define ratio between the SSC luminosity and the synchrotron
luminosity as the $X$ parameter similar to Sari \& Esin (2001), i.e.,
\begin{eqnarray}
 X \equiv \frac{L_{\rm IC}}{L_{\rm syn}}=\frac{U_{\rm ph}}{U_{\rm B}},
\end{eqnarray}
where $U_{\rm ph}$ and $U_{\rm B}$ are the synchrotron photon energy
density and magnetic field energy density, respectively.

For $\nu_{\rm a} < \nu_{\rm m} <\nu_{\rm c}$ (case I) and $\nu_{\rm
m} < \nu_{\rm a} <\nu_{\rm c}$ (case II), the $\nu f_{\nu}$ peaks of
the synchrotron and the SSC components are at $\nu_{\rm c}$ and
$\nu_{\rm cc}^{IC}$, respectively (see Figure 1). One can estimate
\begin{eqnarray}
X = \frac{L_{\rm IC}}{L_{\rm
syn}}&\sim&\frac{\nu_{\rm cc}^{\rm{IC}}f_{\nu}^{\rm{IC}}(\nu_{\rm cc}^{\rm{IC}})}{\nu_{\rm c}f_{\nu}(\nu_{\rm c})}\nonumber\\
&\sim&\frac{\nu_{\rm cc}^{\rm{IC}}R \sigma_T n f_{\rm{max}}
x_0\left(\frac{\nu_{\rm cc}^{\rm{IC}}}{\nu_{\rm mm}^{\rm{IC}}}\right)^{\frac{1-p}{2}}}{\nu_{\rm c}f_{\rm{max}}\left(\frac{\nu_{\rm c}}{\nu_{\rm m}}\right)^{\frac{1-p}{2}}}\nonumber\\
&\sim&4x_0^2\sigma_T n R \gamma_{\rm c}^2\left( \frac{\gamma_{\rm
c}}{\gamma_{\rm m}}\right)^{1-p}, \label{Y-slow}
\end{eqnarray}
which is consistent with Sari \& Esin (2001). Note that when
calculating $X$, we did not include the coefficients in the
analytical approximations of the SSC component, which is of order unity.

For $\nu_{\rm a} < \nu_{\rm c} <\nu_{\rm m}$ (case III), the $\nu f_{\nu}$ peaks of
the synchrotron and SSC components are at $\nu_{\rm m}$, and
$\nu_{\rm mm}^{\rm IC}$, respectively. One therefore has
\begin{eqnarray}
X=\frac{L_{\rm IC}}{L_{\rm
syn}}&\sim&\frac{\nu_{\rm mm}^{\rm{IC}}f_{\nu}^{\rm{IC}}(\nu_{\rm mm}^{\rm{IC}})}{\nu_{\rm m}f_{\nu}(\nu_{\rm m})}\nonumber\\
&\sim&\frac{\nu_{\rm mm}^{\rm{IC}}R \sigma_T n f_{\rm{max}}
x_0\left(\frac{\nu_{\rm mm}^{\rm{IC}}}{\nu_{\rm cc}^{\rm{IC}}}\right)^{-\frac{1}{2}}}{\nu_{\rm m}f_{\rm{max}}\left(\frac{\nu_{\rm m}}{\nu_{\rm c}}\right)^{-\frac{1}{2}}}\nonumber\\
&\sim&4x_0^2\sigma_T n R \gamma_{\rm c}\gamma_{\rm m},
\label{Y-fast}
\end{eqnarray}
which is also consistent with Sari \& Esin (2001).

\section{Strong Synchrotron Self-Absorption Cases}

When $\nu_{\rm a}>\nu_{\rm c}$, synchrotron/SSC cooling and
self-absorption heating would reach a balance around a specific
electron energy under certain conditions (see details in Appendix
A). For such cases, the electron energy distribution and the photon
spectrum are coupled, a numerical iterative procedure is needed to
obtain the self-consistent solution. \cite{Ghisellini88} solved the
kinetic equation and found that the electron energy distribution
would include two components: a thermal component shaped by
synchrotron self-absorption heating, and a non-thermal power-law
component. Based on their results \citep{Ghisellini88}, the electron
distribution is close but not strictly Maxwellian. Strictly, one
needs to use equation (\ref{fc}) to calculate the SSC spectral
component numerically. In the following, we make an approximation to
derive analytical results. For the quasi-thermal component, we take
$N(\gamma)\propto\gamma^{2}$ for $\gamma<\gamma_{\rm a}$ to denote
the thermal component, and take a sharp cutoff at $\gamma_{\rm a}$.
Above this energy, the electron energy distribution is taken as the
standard (broken) power law distribution.

In particular,
for $\nu_{\rm c} < \nu_{\rm a} < \nu_{\rm m}$, the electron distribution becomes
\begin{eqnarray}
\label{fnu1} N(\gamma) = \left\{ \begin{array}{ll}
n\frac{3\gamma^2}{\gamma_{\rm a}^3}, &  \gamma \leq \gamma_{\rm a}, \\
n\gamma_{\rm c}\gamma^{-2}, &  \gamma_{\rm a}<\gamma \leq \gamma_{\rm m}.\\
n\gamma_{\rm m}^{p-1}\gamma_{\rm c}\gamma^{-p-1}, &  \gamma >
\gamma_{\rm m}.
\end{array} \right.
\end{eqnarray}

For $\nu_{\rm m} < \nu_{\rm c} < \nu_{\rm a}$, one has
\begin{eqnarray}
\label{fnu1} N(\gamma) = \left\{ \begin{array}{ll}
n\frac{3\gamma^2}{\gamma_{\rm a}^3}, &  \gamma \leq \gamma_{\rm a}, \\
n(p-1)\gamma_{\rm m}^{p-1}\gamma_{\rm c}\gamma^{-p-1}, &  \gamma >
\gamma_{\rm a}.
\end{array} \right.
\end{eqnarray}
For $\nu_{\rm c} < \nu_{\rm m} < \nu_{\rm a}$, one has
\begin{eqnarray}
\label{fnu1} N(\gamma) = \left\{ \begin{array}{ll}
n\frac{3\gamma^2}{\gamma_{\rm a}^3}, &  \gamma \leq \gamma_{\rm a}, \\
n\gamma_{\rm m}^{p-1}\gamma_{\rm c}\gamma^{-p-1}, &  \gamma >
\gamma_{\rm a}.
\end{array} \right.
\end{eqnarray}

Following these new shapes of the electron distribution, the
synchrotron photon spectra can be calculated, which also contain a
thermal component and a (broken) power-law component. Still applying
equation (\ref{fc}), one can analytically calculate the SSC spectral
component for another three cases in this regime. We note that due
to the simple approximation to the complicated electron pile-up
process, the analytical results presented below are not as precise
as those in the weak absorption cases.

\subsection{Case IV: $\nu_{\rm c} < \nu_{\rm a} < \nu_{\rm m}$}

In this case, the synchrotron photon spectrum reads
\begin{eqnarray}
\label{fnu3} f_{\nu} = \left\{ \begin{array}{ll}
f_{\rm{max}}\left(\frac{\nu}{\nu_{\rm a}}\right)^{2}, &
\nu \leq \nu_{\rm a}; \\
f_{\rm{max}}\mathfrak{R}\left(\frac{\nu}{\nu_{\rm
a}}\right)^{-\frac{1}{2}}, &
\nu_{\rm a}<\nu \leq \nu_{\rm m}; \\
f_{\rm{max}}\mathfrak{R}\left(\frac{\nu_{\rm m}}{\nu_{\rm
a}}\right)^{-\frac{1}{2}}\left(\frac{\nu}{\nu_{\rm
m}}\right)^{-\frac{p}{2}}, & \nu > \nu_{\rm m};
\end{array} \right.
\end{eqnarray}
where $\mathfrak{R}$ is the discontinuity ratio in the electron
distribution at $\gamma_{\rm a}$,
\begin{eqnarray}
\mathfrak{R}=\frac{\gamma_{\rm c}}{3\gamma_{\rm a}}~.
\end{eqnarray}

One can then derive
\begin{eqnarray}
\label{int2fc} I = \left\{ \begin{array}{ll} I_1 \simeq f_{\rm{max}}
x_0 \left(\frac{1}{2}\mathfrak{R}+1\right)
\left(\frac{\nu}{4 \g^2 \nu_{\rm a} x_0}\right), & \nu < 4 \g^2 \nu_{\rm a} x_0 \\
I_2 \simeq \frac{1}{2} f_{\rm{max}} x_0
\mathfrak{R}\left(\frac{\nu}{4 \g^2 \nu_{\rm a}
x_0}\right)^{-\frac{1}{2}}, &
4 \g^2 \nu_{\rm a} x_0 < \nu < 4 \g^2 \nu_{\rm m} x_0 \\
I_3 \simeq \frac{3}{2(p+2)} f_{\rm{max}} x_0\mathfrak{R}
\left(\frac{\nu_{\rm a}}{\nu_{\rm
m}}\right)^{\frac{1}{2}}\left(\frac{\nu}{4 \g^2 \nu_{\rm m}
x_0}\right)^{-\frac{p}{2}}, &
\nu > 4 \g^2 \nu_{\rm m} x_0 \\
\end{array} \right.
\end{eqnarray}
and
\begin{eqnarray}
\label{fc2fc} f_{\nu}^{\rm{IC}} &\simeq& R \sigma_{\rm T} n f_{\rm{max}} x_0 \\
\nonumber &\times&\left\{ \begin{array}{ll}

\left(\frac{1}{2}\mathfrak{R}+1\right)\left(\mathfrak{R}+4\right)
\left(\frac{\nu}{\nu_{\rm aa}^{\rm{IC}}}\right), &
\nu < \nu_{\rm aa}^{\rm{IC}}; \\

\mathfrak{R}\left(\frac{\nu}{\nu_{\rm
aa}^{\rm{IC}}}\right)^{-\frac{1}{2}}
\left[\frac{1}{6}\mathfrak{R}+\frac{9}{10}+\frac{1}{4}\mathfrak{R}\ln{\left(\frac{\nu}{\nu_{\rm
aa}^{\rm{IC}}}\right)}\right], &
\nu_{\rm aa}^{\rm{IC}} < \nu < \nu_{\rm am}^{\rm{IC}}; \\

\mathfrak{R}^2 \left(\frac{\nu}{\nu_{\rm
aa}^{\rm{IC}}}\right)^{-\frac{1}{2}}\left[\frac{3}{p-1} -\frac{1}{2}
+ \frac{3}{4}\ln{\left(\frac{\nu_{\rm mm}^{\rm
IC}}{\nu}\right)}\right], &
\nu_{\rm am}^{\rm{IC}} < \nu < \nu_{\rm mm}^{\rm{IC}}; \\

\frac{9}{2(p+2)}\mathfrak{R}^2 \left(\frac{\nu_{\rm a}}{\nu_{\rm
m}}\right)\left(\frac{\nu}{\nu_{\rm
mm}^{\rm{IC}}}\right)^{-\frac{p}{2}}
\left[\frac{4}{p+3}\left(\frac{\gamma_{\rm a}}{\gamma_{\rm
m}}\right)^{p-1}\frac{\gamma_{\rm a}}{\gamma_{\rm
c}}+\frac{3(p+1)}{(p-1)(p+2)} +\frac{1}{2}\ln{\frac{\nu}{\nu_{\rm
mm}^{\rm IC}}}\right], & \nu
> \nu_{\rm mm}^{\rm{IC}}.
\end{array} \right.
\end{eqnarray}
In this case, there are two peaks in the $\nu F_\nu$ spectrum for
the synchrotron and SSC components, respectively. For the synchrotron
component, the thermal peak is at $(25/9) \nu_{\rm a} \simeq
2.8\nu_a$, and the non-thermal peak is at $\nu_{\rm m}$. For the SSC
component, the thermal peak at $\nu_{\rm aa}^{IC}$, and the
non-thermal peak at $\nu_{\rm mm}^{\rm IC}$. The relative importance
of the two peaks depend on the relative location of $\nu_{\rm a}$
with respect to $\nu_{\rm c}$ and $\nu_{\rm m}$. More specifically,
the spectrum is non-thermal-dominated when $\nu_{\rm a} <
\sqrt{\nu_{\rm m} \nu_{\rm c}}$, and is thermal-dominated when
$\nu_{\rm a} > \sqrt{\nu_{\rm m} \nu_{\rm c}}$.

In Figure 2, we compare the above simplified analytical
approximation (solid) with a simplest power law prescription
(dashed) of the SSC component. The non-thermal-dominated and the
thermal-dominated cases are presented in Figures 2a and 2b,
respectively. Below $\nu_{\rm mm}^{\rm IC}$, similar to the weak
self-absorption regime (cases I-III), the logarithmic terms make the
analytical spectrum harder than the simple broken power-law
approximation above the non-thermal $\nu F_\nu$ peak frequency. At
high frequencies, the simple broken power-law approximation is not
adequate to represent the true SSC spectrum.

\subsection{Case V and VI: $\nu_{\rm a} > {\rm max} (\nu_{\rm m}, \nu_{\rm c})$}

For these two cases ($\nu_{\rm m} < \nu_{\rm c} < \nu_{\rm a}$ and
$\nu_{\rm c} < \nu_{\rm m} < \nu_{\rm a}$), the treatments and results are
rather similar to each other. we take $\nu_{\rm m} < \nu_{\rm c} < \nu_{\rm a}$
as an example. In this case, the synchrotron spectrum reads
\begin{eqnarray}
\label{fnu3} f_{\nu} = \left\{ \begin{array}{ll}
f_{\rm{max}}\left(\frac{\nu}{\nu_{\rm a}}\right)^{2}, &
\nu \leq \nu_{\rm a}; \\
f_{\rm{max}}\mathfrak{R}\left(\frac{\nu}{\nu_{\rm
a}}\right)^{-\frac{p}{2}}, &
\nu > \nu_{\rm a}; \\
\end{array} \right.
\end{eqnarray}
where
\begin{eqnarray}
\mathfrak{R}=(p-1)\frac{\gamma_{\rm c}}{3\gamma_{\rm
a}}\left(\frac{\gamma_{\rm m}}{\gamma_{\rm a}}\right)^{p-1}.
\end{eqnarray}

Applying equation (\ref{fc}), the inner integral $I$ can be then
approximated as
\begin{eqnarray}
\label{int3} I = \left\{ \begin{array}{ll} I_1 \simeq f_{\rm{max}}
x_0\left(\frac{3\mathfrak{R}}{2(p+2)}+1\right)
\left(\frac{\nu}{4 \g^2 \nu_{\rm a} x_0}\right), & \nu < 4 \g^2 \nu_{\rm a} x_0 \\
I_2 \simeq \frac{3}{2(p+2)} f_{\rm{max}} x_0
\mathfrak{R}\left(\frac{\nu}{4 \g^2 \nu_{\rm a}
x_0}\right)^{-\frac{p}{2}}, & \nu > 4 \g^2 \nu_{\rm a} x_0.
\end{array} \right.
\end{eqnarray}

Integrating over the outer integral, one gets
\begin{eqnarray}
\label{fc3} f_{\nu}^{\rm{IC}} &\simeq& R \sigma_{\rm T} n f_{\rm{max}} x_0 \\
\nonumber &\times&\left\{ \begin{array}{ll}
\left(\frac{\mathfrak{3R}}{2(p+2)}+1\right)\left(\frac{\mathfrak{3R}}{p+2}+4\right)
\left(\frac{\nu}{\nu_{\rm aa}^{\rm{IC}}}\right), &
\nu < \nu_{\rm aa}^{\rm{IC}}; \\
\frac{1}{p+2}\left[\frac{6\mathfrak{R}}{p+3}+\mathfrak{R}\left(\frac{9\mathfrak{R}}{2(p+2)}+1\right)+\frac{9\mathfrak{R}^2}{4}\ln{\left(\frac
{\nu}{\nu_{\rm aa}^{\rm{IC}}}\right)}\right]\left(\frac{\nu}{\nu_{\rm aa}^{\rm{IC}}}\right)^{-\frac{p}{2}}, & \nu > \nu_{\rm aa}^{\rm{IC}}; \\
\end{array} \right.
\end{eqnarray}

The case of $\nu_{\rm c} < \nu_{\rm m} < \nu_{\rm a}$ is almost
identical to the above  $\nu_{\rm m} < \nu_{\rm c} < \nu_{\rm a}$.
The only difference is that the expression of $\mathfrak{R}$ is modified
to
\begin{eqnarray}
\mathfrak{R}=\frac{\gamma_{\rm c}}{3\gamma_{\rm
a}}\left(\frac{\gamma_{\rm m}}{\gamma_{\rm a}}\right)^{p-1}.
\end{eqnarray}
This is reasonable, since in the fast cooling case, the electron
energy spectral index is $p=2$, so that the factor $(p-1)$ can be
reduced to 1. The analytical results and simple broken power-law
approximation in this regime is identical to Figure 3c. We note
again that full numerical calculations are needed to obtain
more accurate results.

\begin{figure}
\begin{minipage}[b]{1.0\textwidth}
\centering
\includegraphics[height=3in,width=5in]{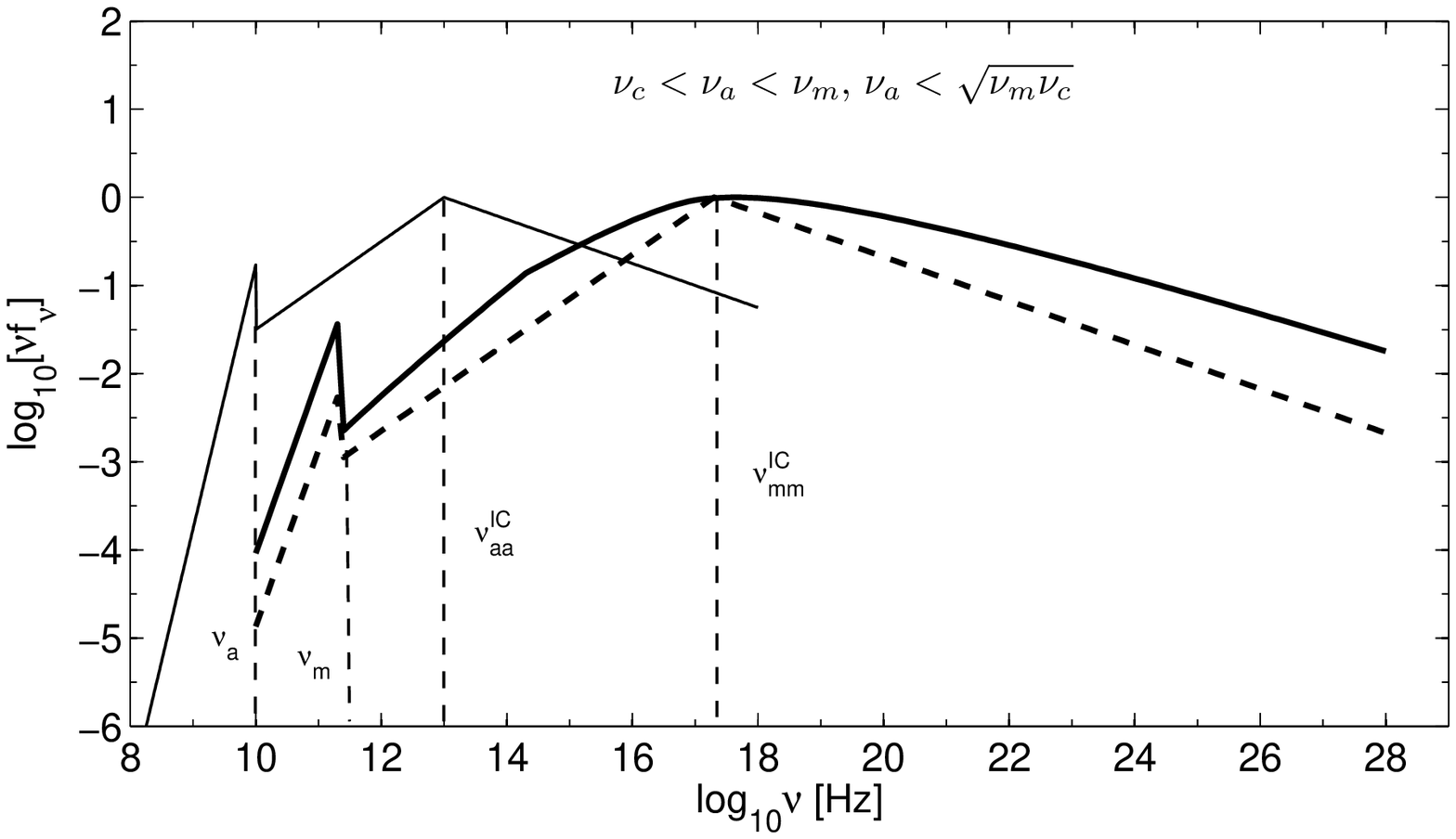}
\end{minipage} \\%
\begin{minipage}[b]{1.0\textwidth}
\centering
\includegraphics[height=3in,width=5in]{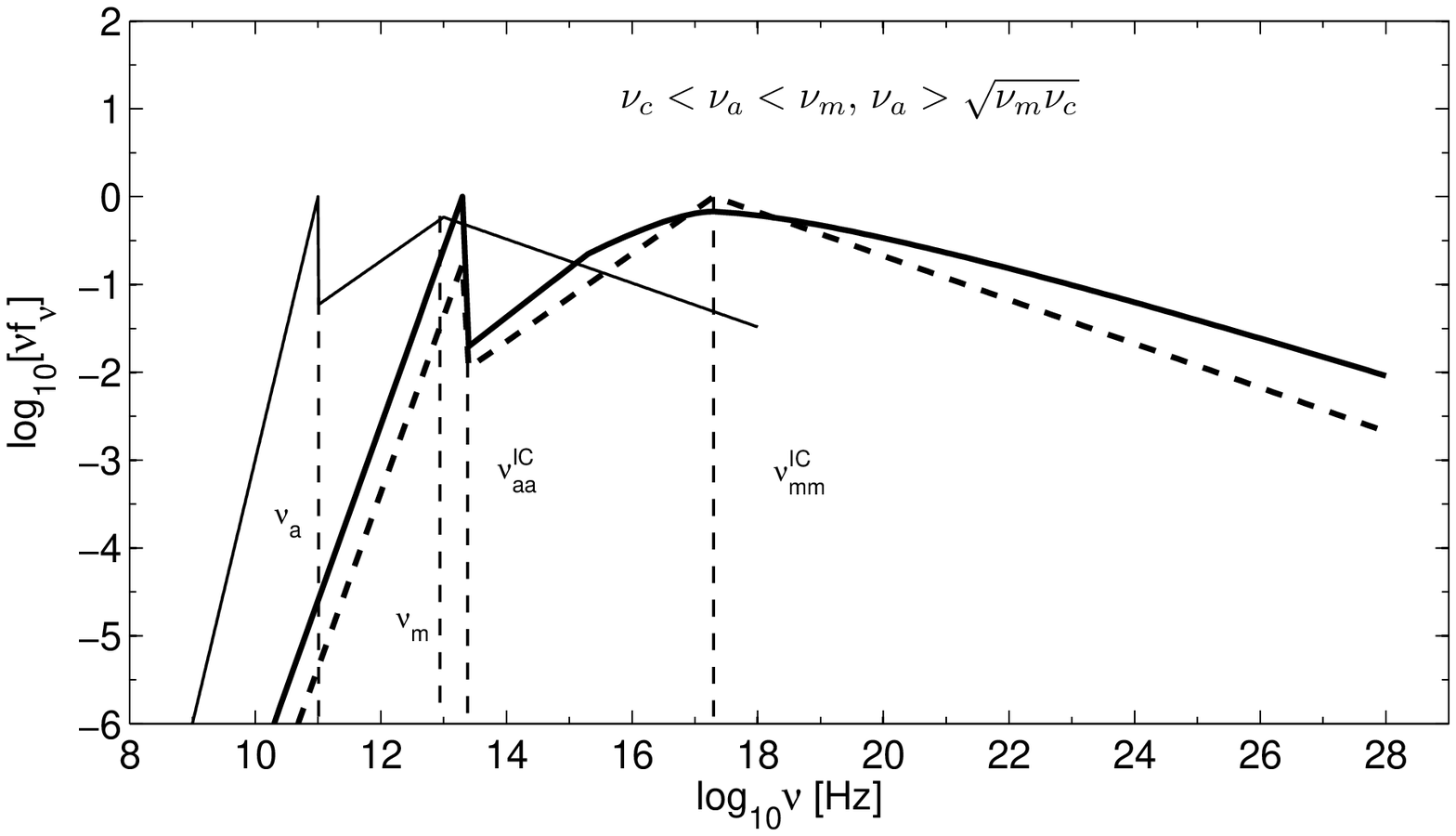}
\end{minipage} \\%
\begin{minipage}[b]{1.0\textwidth}
\centering
\includegraphics[height=3in,width=5in]{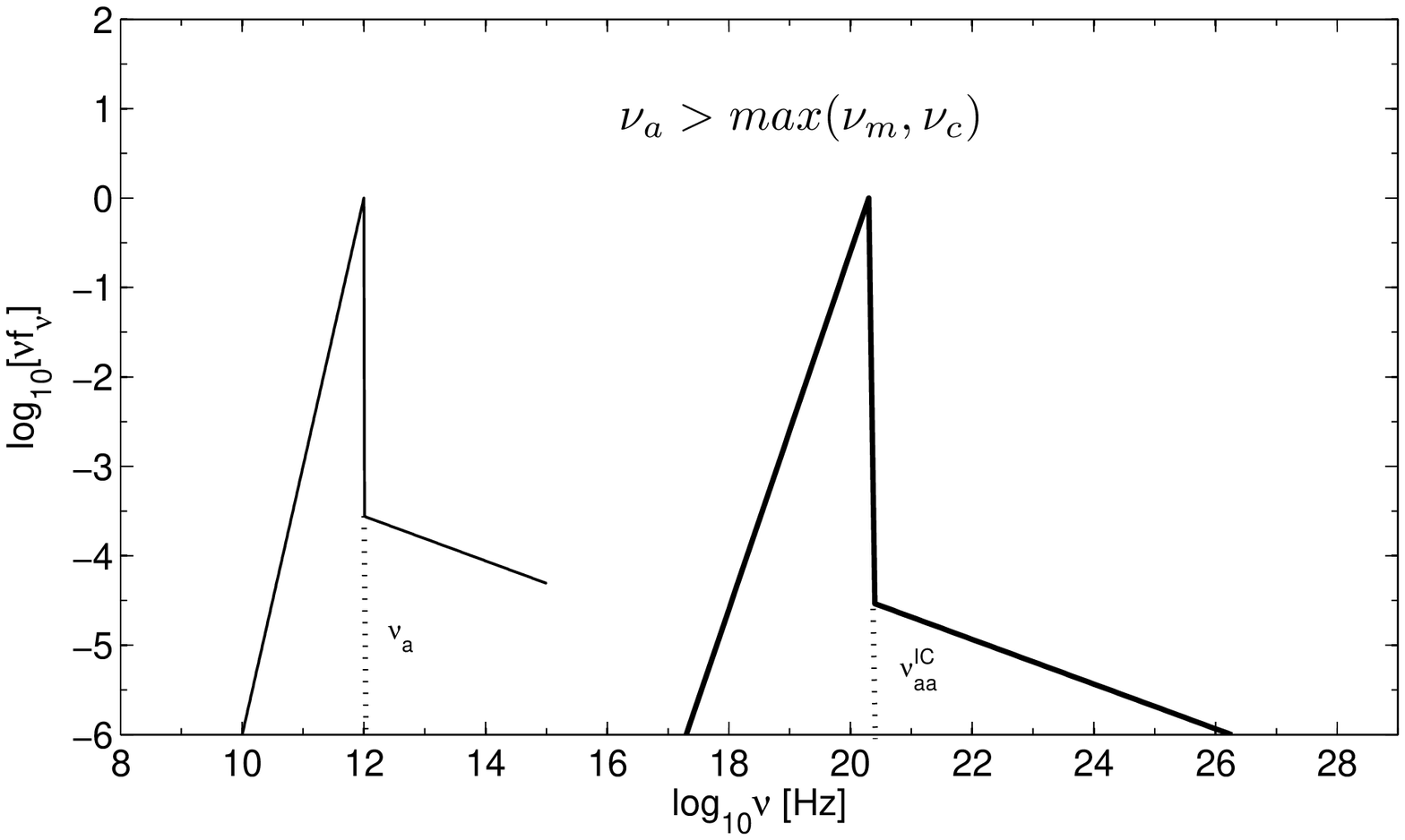}
\end{minipage}%
       \caption{Same as Figure \ref{Fig1}, but for strong synchrotron reabsorption cases .
The top panel is for $\nu_{\rm c} < \nu_{\rm a} <\nu_{\rm m}$ and
$\nu_{\rm a} < \sqrt{\nu_{\rm m} \nu_{\rm c}}$ case; the middle
panel is for $\nu_{\rm c} < \nu_{\rm a} <\nu_{\rm m}$ and $\nu_{\rm
a} > \sqrt{\nu_{\rm m} \nu_{\rm c}}$ case; and the bottom panel is
for $\nu_{\rm a} > {\rm max} (\nu_{\rm m}, \nu_{\rm c})$ case. All
the solid lines are analytical approximations and the dashed lines
are broken power-law approximations.}
           \label{Fig2}
            \end{figure}

Finally, we investigate the $X$ parameter in the strong
synchrotron self-absorption regime.

For $\nu_{\rm c} < \nu_{\rm a} <\nu_{\rm m}$ (case IV), if the
spectrum is non-thermal-dominated, the synchrotron and SSC emission
components peak at $\nu_{\rm m}$ and $\nu_{\rm mm}^{\rm IC}$,
respectively. One thus has
\begin{eqnarray}
X=\frac{L_{\rm IC}}{L_{\rm
syn}}&\sim&\frac{\nu_{\rm mm}^{\rm{IC}}f_{\nu}^{\rm{IC}}(\nu_{\rm mm}^{\rm{IC}})}{\nu_{\rm m}f_{\nu}(\nu_{\rm m})}\nonumber\\
&\sim&\frac{\nu_{\rm mm}^{\rm{IC}}R \sigma_{\rm T} n f_{\rm{max}}
x_0\mathfrak{R}^2\left(\frac{\nu_{\rm mm}^{\rm{IC}}}{\nu_{\rm aa}^{\rm{IC}}}\right)^{-\frac{1}{2}}}{\nu_{\rm m}\mathfrak{R}f_{\rm{max}}\left(\frac{\nu_m}{\nu_a}\right)^{-\frac{1}{2}}}\nonumber\\
&\sim&4x_0^2\sigma_{\rm T} n R \gamma_{\rm c}\gamma_{\rm m}.
\end{eqnarray}
If the spectrum is thermal-dominated, the synchrotron and SSC
emission components peak at $\nu_{\rm a}$ and $\nu_{\rm aa}^{\rm
IC}$, respectively. One has
\begin{eqnarray}
X=\frac{L_{\rm IC}}{L_{\rm
syn}}&\sim&\frac{\nu_{\rm aa}^{\rm{IC}}f_{\nu}^{\rm{IC}}(\nu_{\rm aa}^{\rm{IC}})}{\nu_{\rm a}f_{\nu}(\nu_{\rm a})}\nonumber\\
&\sim&\frac{\nu_{\rm aa}^{\rm{IC}}R \sigma_{\rm T} n f_{\rm{max}} x_0}
{\nu_{\rm a}f_{\rm{max}}}\nonumber\\
&\sim&4x_0^2\sigma_{\rm T} n R \gamma_{\rm a}^2.
\end{eqnarray}
In general, the $X$ parameter for $\nu_{\rm c} < \nu_{\rm a}
<\nu_{\rm m}$ (case IV) is $4x_0^2\sigma_{\rm T} n R \cdot{\rm
max}(\gamma_{\rm a}^2,\gamma_{\rm c}\gamma_{\rm m})$.

For $\nu_{\rm m} < \nu_{\rm c} <\nu_{\rm a}$ (case V) and
$\nu_{\rm c} < \nu_{\rm m} <\nu_{\rm a}$ (case
VI), the synchrotron and SSC emission components peak at $\nu_{\rm a}$ and
$\nu_{\rm aa}^{IC}$, respectively. In this case, one has
\begin{eqnarray}
\label{X6} X=\frac{L_{\rm IC}}{L_{\rm
syn}}&\sim&\frac{\nu_{\rm aa}^{\rm{IC}}f_{\nu}^{\rm{IC}}(\nu_{\rm aa}^{\rm{IC}})}{\nu_{\rm a}f_{\nu}(\nu_{\rm a})}\nonumber\\
&\sim&\frac{\nu_{\rm aa}^{\rm{IC}}R \sigma_{\rm T} n f_{\rm{max}} x_0}
{\nu_{\rm a}f_{\rm{max}}}\nonumber\\
&\sim&4x_0^2\sigma_{\rm T} n R \gamma_{\rm a}^2.
\end{eqnarray}
which is same as the thermal-dominated case for $\nu_{\rm c} <
\nu_{\rm a} <\nu_{\rm m}$ (case IV). So in general the
expression of $X$ is equation (\ref{X6}) only if the spectrum is
thermal-dominated.

\section{Conclusion and discussion}

We have extended the analysis of \cite{Sari01} and derived the
analytical approximations of the SSC spectra of all possible orders
of the three synchrotron characteristic frequencies
$\nu_{\rm a}$, $\nu_{\rm m}$, and $\nu_{\rm c}$. Based on the relative
order between $\nu_{\rm a}$ and $\nu_{\rm c}$, we divide the six possible
orders into two regimes.

In the weak self-absorption regime
$\nu_{\rm a}<\nu_{\rm c}$, self-absorption does not affect the
electron energy distribution. Two cases in this regime have been
studied by \cite{Sari01}. Our results are consistent with theirs
(except the two typos in their paper).
For the other regime $\nu_{\rm m}<\nu_{\rm a}<\nu_{\rm c}$,
we find that the SSC spectrum is linear to $\nu$ all the way to
$\nu_{\rm ma}^{\rm{IC}}$,
and there is no break corresponding to $\nu_{\rm mm}^{\rm{IC}}$.

In the strong self-absorption $\nu_{\rm a}>\nu_{\rm c}$ regime,
synchrotron self-absorption heating balances synchrotron/SSC
cooling, leading to pile-up of electrons at a certain energy, so
that the electron energy distribution is significantly altered, with
an additional thermal component besides the non-thermal power law
component. Both the synchrotron and the SSC spectral components
become two-hump shaped. To get an analytical approximation of the
SSC spectrum, we simplified the quasi-thermal electron energy
distribution as a power law with a sharp
cutoff above the piling up energy, and derived the analytical
approximation results of the synchrotron and SSC spectral components.
We suggest that for the
thermal-dominated cases, i.e. $\nu_{\rm a} > \sqrt{\nu_{\rm
m}\nu_{\rm c}}$ in the $\nu_{\rm c} < \nu_{\rm a} < \nu_{\rm m}$
regime or the $\nu_{\rm a}
> \max (\nu_{\rm m}, \nu_{\rm c})$ regime, full numerical
calculations are needed to get accurate results.

In general, the SSC component roughly tracks the shape of the seed
synchrotron component, but is smoother and harder at high energies.
For all the cases, we compare our analytical approximation results
of SSC component with the simplest broken power-law prescription. We
find that in general the presence of the logarithmic terms in the
high energy range makes the SSC spectrum harder than the broken
power-law approximation. One should consider these terms when
studying high energy emission. The only exceptions are the $\nu_{\rm
a} > \max (\nu_{\rm m}, \nu_{\rm c})$ regimes. However, in these
regimes the analytical approximations may be no longer good, and one
should appeal to full numerical calculations.

Our newly derived spectral regimes may find applications in
astrophysical objects with high ``compactness'' (i.e. high luminosity,
and small size). In these cases, $\nu_{\rm a}$ can
be higher than $\nu_{\rm c}$ or $\nu_{\rm m}$, or even both
(see Appendix A for the critical condition).
For example, in the
early afterglow phase of GRBs, slow cooling may be relevant, and the
radio afterglow is self-absorbed with $\nu_{\rm a}$ above $\nu_{\rm m}$
\citep[e.g.][]{Chandra12}.  In the prompt emission phase when fast
cooling is more relevant, the self-absorption frequency can be above
$\nu_{\rm c}$ \cite[e.g.][]{Shen09}.

An example of the extreme case $\nu_{\rm a} > {\rm max} (\nu_{\rm
m}, \nu_{\rm c})$ can be identified for a GRB problem. For a dense
circumburst medium with a wind-like ($n \propto r^{-2}$) structure,
in the reverse shock region, the condition $\nu_{\rm a} > \max
(\nu_{\rm m}, \nu_{\rm c})$ can be satisfied
\citep[e.g.][]{Kobayashi04}. For a GRB with isotropic energy
$E=10^{52} E_{52}$, initial Lorentz factor $\Gamma_0=100
\Gamma_{0,2}$, initial shell width $\Delta = 10^{12} \Delta_{12}$
running into stellar wind with density $\rho = (5\times 10^{11} {\rm
g~cm^{-1}}) A_* r^{-2}$, one can derive following parameters at the
shock crossing radius $r_\times$: The blastwave Lorentz factor
$\Gamma_\times = 25.8 A_*^{-1/4} \Delta_{12}^{-1/4} E_{52}^{1/4}$,
$\nu_{\rm m} = 3.1 \times 10^{14} ~{\rm Hz}~ [g(p)/g(2.3)] A_*
\Delta_{12}^{-1/2} E_{52}^{-1/2} \epsilon_{e,-1}^2
\epsilon_{B,-2}^{1/2} \Gamma_{0,2}^2$, $\nu_c = 1.2\times 10^{12}
~{\rm Hz}~ A_*^{-2} \Delta_{12}^{1/2} E_{52}^{1/2}
\epsilon_{B,-2}^{-3/2}$, $\nu_a = 4.6\times 10^{14}~{\rm Hz}~
A_*^{3/5} \Delta_{12}^{-11/10} E_{52}^{1/10} \epsilon_{B.-2}^{3/10}
\Gamma_{0,2}^{-2/5}$. Here $\epsilon_e = 0.1 \epsilon_{e,-1}$ and
$\epsilon_B = 0.01 \epsilon_{B,-2}$ are microphysics shock
parameters for the internal energy fraction that goes to electrons
and magnetic fields, $p$ is the electron spectral index, and $g(p) =
(p-2)/(p-1)$. We can see that for typical parameters, $\nu_a > {\rm
max} (\nu_c, \nu_m)$ is satisfied. In this regime, one should check
whether the ``Razin'' plasma effect is important. At shock crossing
time, the comoving number density of the shocked ejecta region is
$n' = 2.3 \times 10^8 ~{\rm cm^{-3}}~ A_*^{5/4} \Delta_{12}^{-7/4}
E_{52}^{-1/4} \Gamma_{0,1}^{-1}$. Noticing that the comoving plasma
angular frequency is $\omega'_p = 5.63 \times 10^4 ~{\rm s^{-1}}
{n'}^{1/2}$, one can write the plasma frequency in the observer
frame as $\nu_p = 1.4\times 10^{11}~{\rm Hz}~ A_*^{3/8}
\Delta_{12}^{-9/8} E_{52}^{1/8} \Gamma_{0,2}^{-1/2}$. Multiplying by
$\gamma_{\rm a} =102 A_*^{1/20} \Delta_{12}^{-1/20} E_{52}^{1/20}
\epsilon_{B.-2}^{-1/10} \Gamma_{0,2}^{-1/5}$, one gets $\gamma_{\rm
a}\nu_p = 1.4 \times 10^{13}~{\rm Hz}~ A_*^{17/40}
\Delta_{12}^{-47/40} E_{52}^{7/40} \epsilon_{B,-2}^{-1/10}
\Gamma_{0,2}^{-7/10}$, which is much smaller than $\nu_{\rm a}$.
This suggests that the Razin effect is not important \citep{ryl79},
and the dominant mechanism to suppress synchrotron emission at low
energies is synchrotron self-absorption. Notice that for this
particular problem, the second order Comptonization may not be
suppressed by the Klein-Nishina effect, and one has to introduce
it for a fully self-consistent treatment.

\section*{Acknowledgements}

We thank the referee for constructive comments, Martin Rees for an
important remark, and Zhuo Li, Resmi Lekshmi and Yuan-Chuan Zou for
helpful discussions. We acknowledge the National Basic Research
Program (``973" Program) of China under Grant No. 2009CB824800. This
work is supported by NSF under Grant No. AST-0908362. WHL
acknowledges support by National Natural Science Foundation of China
(grants 11003004, 11173011 and U1231101), HG and WHL acknowledge
Fellowship support from China Scholarship Program, and XFW
acknowledges support by the One-Hundred-Talents Program.

\appendix
\section{Condition of electron pile-up and strong absorption}

By applying the Einstein coefficients and their relations to a system
with three energy levels, \cite{Ghisellini91} have derived one
useful analytical expression of the cross section for synchrotron
self-absorption:
\begin{eqnarray}
\label{crosectionG} \sigma_{\rm S}(\gamma,\nu) = \left\{
\begin{array}{ll}
\frac{2^{2/3}\sqrt{3}\pi \Gamma^2(4/3)}{5}\frac{\sigma_{\rm T}}{\alpha_{\rm f}}\frac{B_{\rm cr}}{B}\left(\frac{\gamma\nu}{3\nu_{\rm L}}\right)^{-5/3}, &  \frac{\nu_{\rm L}}{\gamma} < \nu \ll \frac{3}{2}\gamma^2\nu_{\rm L}, \\
\frac{\sqrt{3}}{2}\pi^2\frac{\sigma_{\rm T}}{\alpha_{\rm
f}}\frac{B_{\rm cr}}{B}\frac{1}{\gamma^3}\left(\frac{\nu_{\rm
L}}{\nu}\right)\rm{exp}\left(\frac{-2\nu}{3\gamma^2\nu_{\rm
L}}\right), & \nu \gg \frac{3}{2}\gamma^2\nu_{\rm L}.
\end{array} \right.
\end{eqnarray}
where $\gamma$ is the relevant electron Lorentz factor, $\nu$ is
photon frequency being absorbed, $\alpha_{\rm f}$ is the fine
structure constant, $B_{\rm cr}=\alpha_{\rm
f}(m_ec^2/r_e^3)^{1/2}\approx4.4\times10^{13} \rm G$ is the critical
magnetic field strength, $r_e$ is the classical electron radius, and
$\nu_{\rm L}=eB/2\pi m_ec$ is the electron cyclotron frequency. All
the parameters introduced in this section are in the comoving frame.

For a simple derivation of the electron pile-up condition, we take
an approximate form
\begin{eqnarray}
\label{crosection} \sigma_{\rm S}(\gamma,\nu) = \left\{
\begin{array}{ll}
\frac{2^{2/3}\sqrt{3}\pi \Gamma^2(4/3)}{5}\frac{\sigma_{\rm T}}{\alpha_{\rm f}}\frac{B_{\rm cr}}{B}\left(\frac{\gamma\nu}{3\nu_{\rm L}}\right)^{-5/3}, &  \frac{\nu_{\rm L}}{\gamma} < \nu \leq \frac{3}{2}\gamma^2\nu_{\rm L}, \\
0, & \nu > \frac{3}{2}\gamma^2\nu_{\rm L}.
\end{array} \right.
\end{eqnarray}

For electrons with Lorenz factor $\gamma$, the heating rate due to
synchrotron self-absorption can be estimated as
\begin{eqnarray}
\dot{\gamma}^+(\gamma)=\int_{0}^{\infty}c\cdot n_{\nu}\cdot
h\nu\cdot \sigma_{\rm S}(\gamma,\nu) \cdot d\nu
\end{eqnarray}
where $n_{\nu}$ is the specific photon number density at frequency
$\nu$ contributed by all the electrons.

The cooling rate for electrons with Lorentz factor of $\gamma$ is
\begin{eqnarray}
\dot{\gamma}^-(\gamma)&=&(1+Y)\cdot P_{\rm syn}\nonumber\\
&=&(1+Y)\times\frac{4}{3}\sigma_{\rm T}c\gamma^2\frac{B^2}{8\pi},
\end{eqnarray}
where $Y\equiv\frac{P_{\rm ssc}}{P_{\rm syn}}$ is the Compton
parameter.

By balancing the heating and cooing rate, one can easily obtain
the critical electron Lorentz factor $\gamma_{\rm cr}$, which satisfies
\begin{eqnarray}
\label{cri} \dot{\gamma}^+(\gamma_{\rm cr})=\dot{\gamma}^-(\gamma_{\rm cr})
\end{eqnarray}
Initially, the photon spectrum has not been revised through
self-absorption, i.e., $n_{\nu}\propto \nu^{1/3}$. One therefore has
\begin{eqnarray}
\gamma_{\rm cr} =2.1\times10^4B^{-3/5}{\cal F}_{\rm \nu,
max}^{3/10}\gamma_{c}^{-1/5}(1+Y)^{-3/10}
\end{eqnarray}

The electron pile-up (strong absorption) condition can be expressed
as
\begin{eqnarray}
\label{cri} \gamma_{\rm cr} > \gamma_{c}=\frac{6\pi m_e
c}{\sigma_{\rm T} B^2 t(1+Y)}~.
\end{eqnarray}
With equations \ref{crosection} - \ref{cri}, the pile-up condition
can be expressed as
\begin{eqnarray}
\left(\frac{B}{100\rm G}\right)^2\times\left(\frac{t}{100\rm
s}\right)^{4/3}\times {\cal F}_{\rm \nu,
max}^{1/3}\times\left(\frac{1+Y}{2}\right)>1 \label{criterion}
\end{eqnarray}
where
\begin{eqnarray}
{\cal F}_{\rm \nu, max}=\frac{f_{\rm
max}}{\Gamma(1+z)}\left(\frac{d_{\rm L}}{R}\right)^2=1~{\rm
erg~cm^{-2}~s^{-1}~Hz^{-1}}~\frac{f_{\rm max,
mJy}}{\Gamma_{2}(1+z)}\left(\frac{d_{\rm L,28}}{R_{14}}\right)^2
\end{eqnarray}
is the synchrotron peak flux in the emission region. Here $d_{\rm
L}$ is the luminosity distance of the source, and $R$ is the
distance of the emission region from the central engine.

One can immediately see that this condition is very difficult to
satisfy. It requires a strong magnetic field, long dynamical
time scale and high synchrotron flux. In the GRB afterglow problem,
for forward shock emission, $B$ decreases with $t$ rapidly, and
there is essentially no parameter space to satisfy the condition.
This condition may be realized in extreme conditions, e.g. the
reverse shock emission during shock crossing phase for a wind
medium, as discussed in Sect.4 in the main text.

One interesting note is that SSC cooling only enhances the pile-up
condition. Once the pile-up condition is satisfied for synchrotron
cooling only, adding SSC cooling only makes the condition more
easily satisfied (as shown in equation \ref{criterion}).

Once the electron pile-up process is triggered, both electron
distribution and photon spectrum would be modified, so that the
value of $\gamma_{\rm cr}$ is modified correspondingly. According to
the numerical calculation results
\citep{Ghisellini88,Ghisellini91,Ghisellini98a}, the electron
distribution is dominated by a quasi-thermal component until a
``transition'' Lorentz factor $\gamma_{\rm t}$, above which the
electrons go back to the optically-thin normal power law. In this
case, $\gamma_{\rm cr}$ should be around the thermal peak, and
$\gamma_a$ should be around the ``transition'' Lorentz factor
$\gamma_{\rm t}$, which is slightly larger than $\gamma_{\rm cr}$.
Consequently, one would roughly have $\gamma_a \sim \gamma_{\rm cr}
\sim \gamma_{t}$, so that the assumption of a sharp cutoff in the
electron distribution around this energy is justified. In the main
text, we did not differentiate the three Lorentz factors, and only
adopt $\gamma_a$ in the expressions.

\label{lastpage}

\end{document}